\documentclass{iaus}
\usepackage{graphicx}

\title[A Relativistic Motion Integrator (RMI)] %% give here short title %%
{A Relativistic Motion Integrator:\\ {\Large Numerical accuracy and illustration with BepiColombo and Mars-NEXT}}

\author[A. Hees \and S. Pireaux]   %% give here short author list %%
{A. Hees$^{1}$ \and S. Pireaux$^{2}$}

\affiliation{$^{1,2}$ Royal Observatory of Belgium (ROB) \\ Avenue Circulaire 3, 1180 Bruxelles, Belgium\\[\affilskip]
$^1$ aurelien.hees@oma.be \quad $^2$ sophie.pireaux@oma.be }

\pubyear{2009}
\volume{162}  %% insert here IAU Symposium No.
\jname{Relativity in Fundamental Astronomy: Dynamics, Reference Frames and Data Analysis}
\editors{S. Klioner, P. K. Seidelmann \& M. Soffel}

\DeclareTextSymbol{\degre}{T1}{6}
\DeclareTextSymbol{\degre}{OT1}{23}
\begin{document}
%le symbole degré

\maketitle

\begin{abstract}
 Today, the motion of spacecraft is still described by the classical Newtonian equations of motion plus some relativistic corrections. This approach might become cumbersome due to the increasing precision required. We use the Relativistic Motion Integrator (RMI) approach to numerically integrate the native relativistic equations of motion for a spacecraft. The principle of RMI is presented. We compare the results obtained with the RMI method with those from the usual Newton plus correction approach for the orbit  of the BepiColombo (around Mercury) and Mars-NEXT (around Mars) orbiters. Finally, we present a numerical study of RMI and we show that the RMI approach is relevant to study the orbit of spacecraft.
\end{abstract}

\firstsection % if your document starts with a section,

\section{The Relativistic Motion Integrator (RMI) and an analytical development}
The software RMI presented in~\cite{Pireaux2006,Pireaux2008} numerically integrates the relativistic equations of motion 
\begin{equation}\label{geodesic}
 \frac{d^2X^\alpha}{d\tau^2}=-\Gamma^\alpha_{\mu\nu}\frac{dX^\mu}{d\tau}\frac{dX^\nu}{d\tau}
\end{equation}
for a given metric $G_{\mu\nu}$ where $X^\mu=(cT,X,Y,Z)$ are the coordinates, $\tau$ is the proper time and $\Gamma^\alpha_{\mu\nu}$ are the Christoffel symbols of the metric considered, derived numerically.
As an example, we use this integrator with the planetocentric metric advised by the IAU 2000 resolutions, described in~\cite{Soffel:2003} and characterised by a scalar potential $W$ and a vector potential $W^i$.
Until now, we considered only the central body of mass $M$, so that we have:
\begin{equation}
 \left\{
\begin{array}{rcl}
W(X^\alpha)&=&\frac{GM}{R}\left[ 1+ \sum_{l=2}^\infty \sum_{m=0}^l \left(\frac{R_e}{R}\right)^l P_{lm}(\cos \theta) \left(C_{lm}\cos m\phi + S_{lm} \sin m\phi\right)\right]\\
W^i(X^\alpha)&=&-\frac{G}{2}\frac{\left(\mathbf{R}\times\mathbf{S}\right)^i}{R^3}
\end{array}
\right.
\end{equation}
where $G$ is Newton's gravitational constant, $c$ the speed of light, $C_{lm}$ and $S_{lm}$ are related to the central gravity field, $R_e$ is the equatorial radius of the central body, while the vector $\mathbf{S}$ is its spin moment and $R=\sqrt{X^2+Y^2+Z^2}$.

It is possible to develop analytically the equations of motion (\ref{geodesic}) at first Post-Newtonian (1PN) order. Doing so, one gets $\frac{d^2\mathbf R}{dt^2}=-\frac{GM}{R^3}\mathbf R +\textrm{corr}$, where the corrections are composed of different types of forces: a Newtonian correction coming from the harmonics (proportional to $C_{lm}$ or $S_{lm}$), a relativistic Schwarzschild acceleration (proportional to $1/c^2$), a relativistic correction coming from the harmonics (proportional to $C_{lm}/c^2$ or $S_{lm}/c^2$), a relativistic coupling between harmonics (proportional to $C_{lm}C_{lm}/c^2$, $C_{lm}S_{lm}/c^2$ or $S_{lm}S_{lm}/c^2$) and finally a relativistic Lense-Thirring acceleration (proportional to the spin momentum over $c^2$).

\section{Results for the BepiColombo and Mars-NEXT missions}
\begin{figure}[t]
\begin{minipage}[t]{0.5\linewidth}
\begin{center}
	\includegraphics[height=2.8cm,width=1.\linewidth]{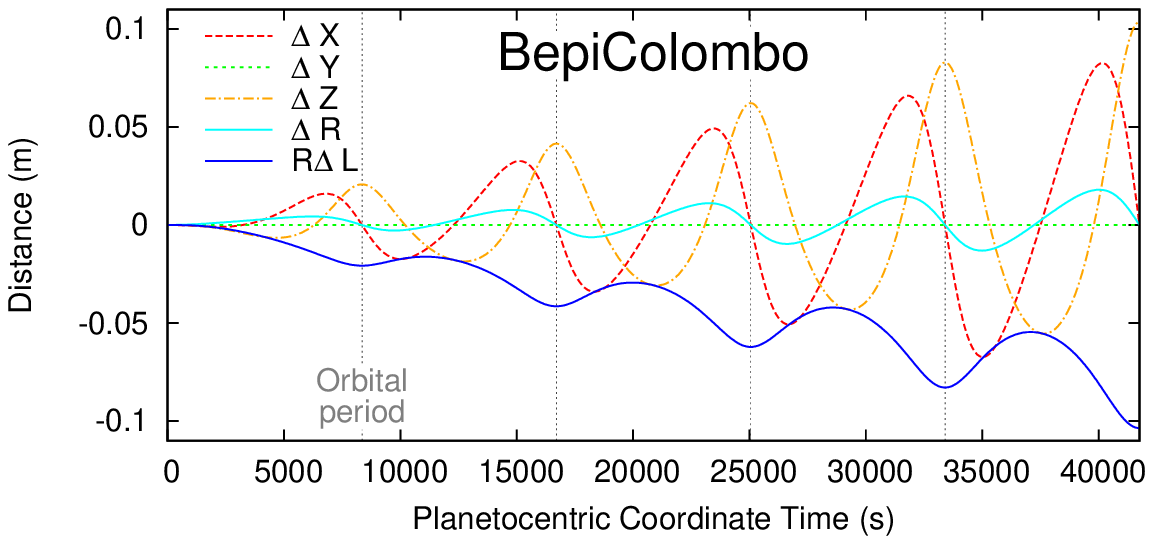}
	\includegraphics[height=2.8cm,width=1.\linewidth]{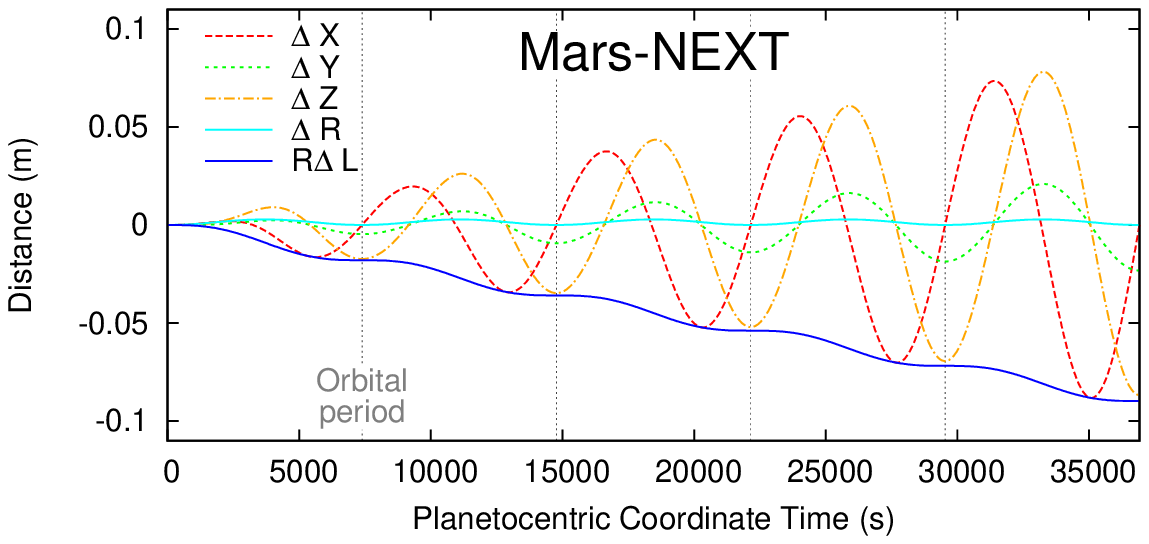}
	\caption{Corrections due to the relativistic Schwarzschild acceleration.}
	\label{figSchwarz}
\end{center}
\end{minipage}
\begin{minipage}[t]{0.5\linewidth}
\begin{center}
	\includegraphics[height=2.8cm,width=1.\linewidth]{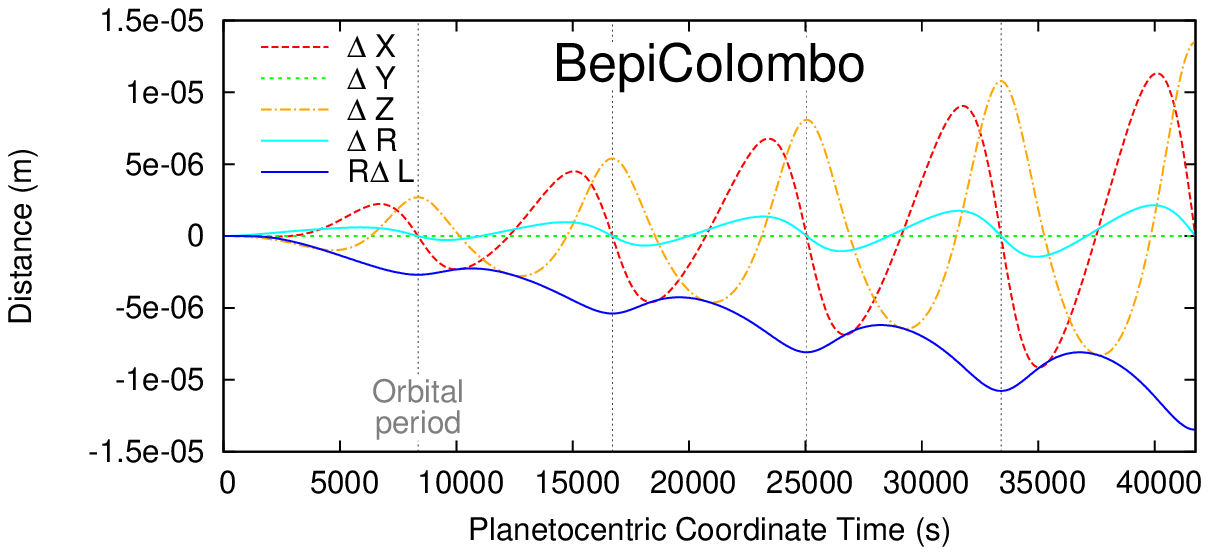}
	\includegraphics[height=2.8cm,width=1.\linewidth]{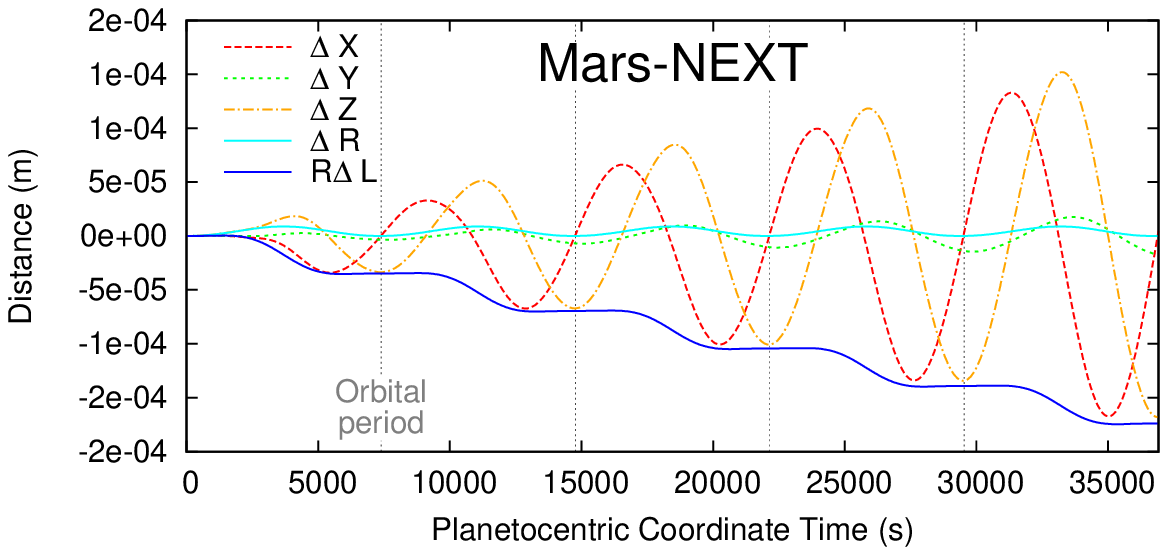}
	\caption{Corrections due to the relativistic contribution from the first harmonics ($C_{20}$, $C_{22}$ and $S_{22}$).}
	\label{figHarm}
\end{center}
\end{minipage}
\end{figure}

The analytical development described above is used to assess the order of magnitude of each separate effect and to validate the RMI method. Figures \ref{figSchwarz}, \ref{figHarm} and \ref{figLense} show the separate impact of the relativistic effects in terms of cartesian coordinates $(X,Y,Z)$, radial distance $R$ and $L=\omega+w$, where $\omega$ is the argument of pericenter and $w$ is the true anomaly. The orbital parameters of the BepiColombo mission can be found in~\cite{Bepi}: $a=3389 \ km$, $e=0.162$, $i=90$\degre. The orbital parameters of the Mars-NEXT mission (see~\cite{MarsNext}) are: $a=3896 \ km$, $e=0$ and $i=75$\degre. The numerical integration has been performed over 5 orbital periods.

\section{Numerical precision}
\begin{figure}
\begin{minipage}{0.5\linewidth}
\begin{center}
 	\includegraphics[height=2.8cm,width=1.0\linewidth]{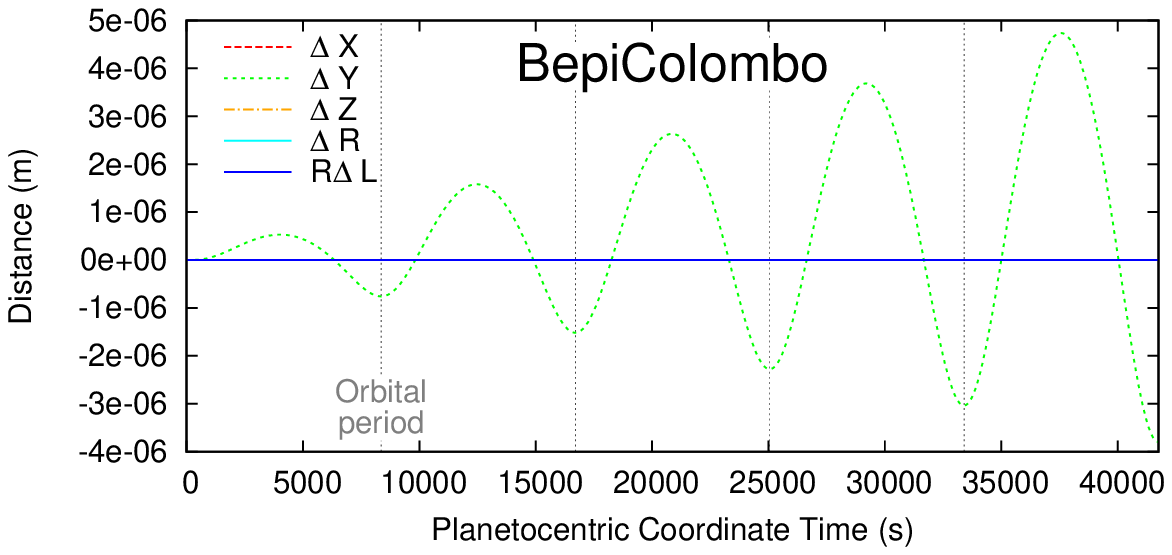}
	\includegraphics[height=2.8cm,width=1.0\linewidth]{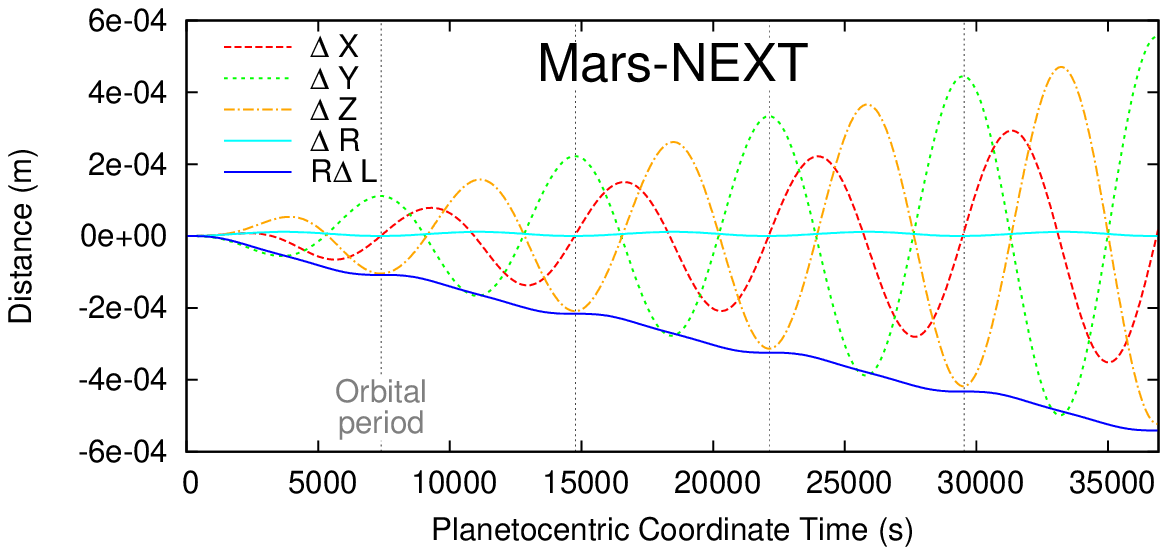}
	\caption{Corrections due to the relativistic Lense-Thirring acceleration.}
	\label{figLense}
\end{center}
\end{minipage}
\begin{minipage}{0.5\linewidth}
\begin{center}

 \includegraphics[width=0.9\textwidth]{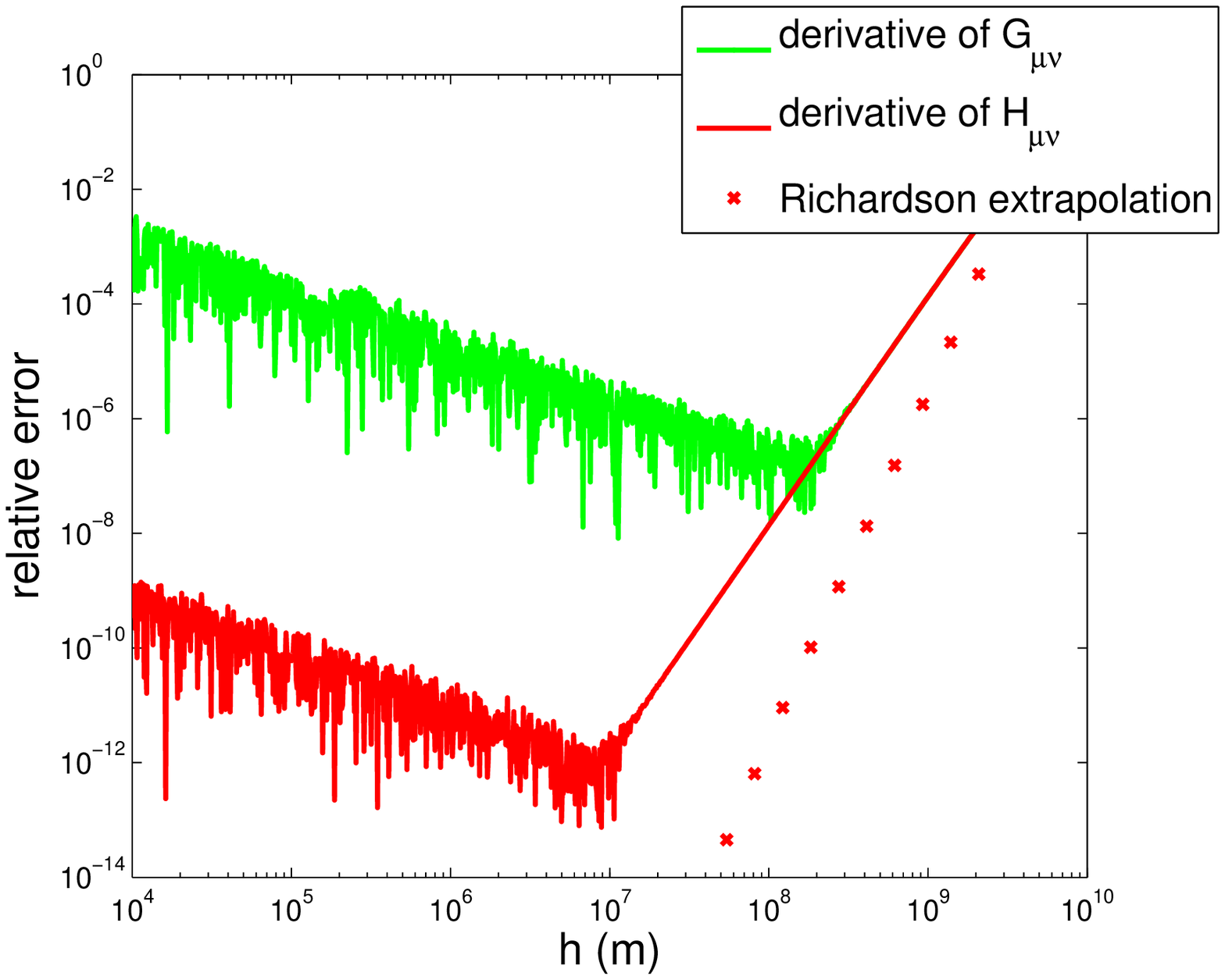}
 % fig1f.: 0x0 pixel, 0dpi, nanxnan cm, bb=
 \caption{Relative precision of the metric derivative as a function of the discretization step size $h$.}
 \label{figDer}
\end{center}
\end{minipage}
\end{figure}

The numerical derivative has to be treated carefully. In the implementation of RMI, we used a fourth-order numerical derivative
\begin{equation}
 f'(x)\approx D_h+\mathcal O(h^4) =\frac{f(x-2h)-8f(x-h)+8f(x+h)-f(x+2h)}{12h}+\mathcal O(h^4).
\end{equation}
For large $h$, the discretization error is important, while for small $h$, the roundoff error increases. We use an optimal derivation step computed analytically for a function $\frac{1}{r}$ (see Figure \ref{figDer}), given by  $h_{\textrm{\footnotesize opt}}=(45\epsilon a^6c^2/(960GM))^{1/5}$ (\cite{numerical}) where $\epsilon$ is the machine precision and $a$ is the semi-major axis. As can be seen on Figure \ref{figDer}, we derive $H_{\mu\nu}=G_{\mu\nu}-\eta_{\mu\nu}$, with $\eta_{\mu\nu}$ the Minkowski metric, instead of $G_{\mu\nu}$ since it is more stable numerically. Moreover, we use a Richardson extrapolation in order to increase the precision on the derivative (\cite{richardson}). This extrapolation uses two estimations of the derivative of order 4 with different step size ($D_h$ and $D_{h/k}$ with $k$ a real factor) to construct an estimation of order 8
\begin{equation}
 f'(x)\approx \frac{k^4D_{h/k}-D_h}{k^4-1}+\mathcal O(h^8).
\end{equation}
Typically, the value of $k$ is often chosen as 2 or 1.5. We can see on Figure \ref{figDer} that this procedure can increase the derivative precision. It is furthermore less sensitive to the choice of $h_{\textrm{\footnotesize opt}}$. With such an implementation, it is possible to show that the relative precision of RMI is of the order of $10^{-12}$ in double precision and $10^{-22}$ in quadruple precision.

\section{Conclusion}
We have shown that RMI is useful to compute relativistic orbits for different missions. It includes all the relativistic effects (up to the corresponding order of the metric). It is quite easy to use, since the user only has to change the metric module if he wants to change the metric. Until now, RMI is only a prototype that is more time consuming than the usual 1PN approach. Nevertheless, it is possible to reduce drastically the integration time required by the RMI method via proper coding and using parallelization (the computation of the Christoffel symbols can easily be parallelized).
% Using this software we have shown that the relativistic corrections for the BepiColombo orbiter and the Mars-NEXT orbiter (considering only the central body) are of the order of 10 $cm$ after 5 orbital periods.

\section*{Acknowledgments}
A.~Hees is a research fellow from the FRS-FNRS (Belgian Fund for Scientific Research) for his PhD thesis at ROB-UCL (Universit\'e Catholique de Louvain, Belgium) and both authors acknowledge a FNRS and a LOC grant to attend to the IAU 261 symposium.

\end{document}